\documentclass[RNAAS]{aastex631}

\begin{document}

\title{\large Overview and Validation of the Asteroseismic Modeling Portal v2.0}

\author[0000-0003-4034-0416]{Travis S.~Metcalfe}
\affiliation{White Dwarf Research Corporation, 9020 Brumm Trail, Golden, CO 80403, USA}

\author[0000-0002-2522-8605]{Richard H.~D.~Townsend}
\affiliation{Department of Astronomy, University of Wisconsin-Madison, Madison, WI 53706, USA}

\author[0000-0002-4773-1017]{Warrick H.~Ball}
\affiliation{School of Physics \& Astronomy, University of Birmingham, Edgbaston, Birmingham B15 2TT, UK}
\affiliation{Stellar Astrophysics Centre, Aarhus University, Ny Munkegade 120, DK-8000 Aarhus C, Denmark}

\begin{abstract} 

The launch of NASA's Kepler space telescope in 2009 revolutionized the quality and 
quantity of observational data available for asteroseismic analysis. While Kepler was 
able to detect solar-like oscillations in hundreds of main-sequence and subgiant stars, 
the Transiting Exoplanet Survey Satellite (TESS) is now making similar observations for 
thousands of the brightest stars in the sky. The Asteroseismic Modeling Portal (AMP) is 
an automated and objective stellar model-fitting pipeline for asteroseismic data, which 
was originally developed to use models from the Aarhus Stellar Evolution Code (ASTEC). We 
briefly summarize an updated version of the AMP pipeline that uses Modules for 
Experiments in Stellar Astrophysics (MESA), and we present initial modeling results for 
the Sun and several solar analogs to validate the precision and accuracy of the inferred 
stellar properties.

\end{abstract}

\section{Overview}

The Asteroseismic Modeling Portal (AMP) was originally released in 2009 
\citep{Metcalfe2009, Woitaszek2009}, and several minor revisions followed as the quality 
of asteroseismic data from the Kepler mission gradually improved \citep{Mathur2012, 
Metcalfe2014, Creevey2017}. The optimization approach used a parallel genetic algorithm 
\citep[GA;][]{Metcalfe2003} to match a given set of observations with models from the 
Aarhus stellar evolution and pulsation codes \citep{JCD2008a, JCD2008b}. Below we briefly 
summarize a major revision of AMP (v2.0), which couples the same optimization method to 
the MESA \citep{Paxton2019} and GYRE \citep{Townsend2013} codes. We focus on differences 
in the underlying physics, updated treatments and methods that are distinct from previous 
versions, and the initial validation using the Sun and several solar analogs.\\

\section{Input Physics and Methods}

The choices of input physics for AMP~2.0 are almost all the defaults in MESA release 
12778. MESA's default equation of state is predominantly a blend of OPAL 
\citep{Rogers2002} and SCVH \citep{Saumon1995}. Opacities are taken from OPAL 
\citep{Iglesias1993, Iglesias1996} at high temperature ($\log T\geq3.88$), from 
\citet{Ferguson2005} at low temperature ($\log T\leq3.80$), and blended smoothly between 
these temperature ranges. The solar mixture is that of \citet{Grevesse1998} and nuclear 
reaction rates are taken from NACRE \citep{Angulo1999}. Convective heat and composition 
transport is described using the mixing-length model by \citet{Cox1968}. The 
mixing-length parameter itself is left free in the optimization. Mixing by convective 
overshooting, when included, is modeled using an exponentially-decaying diffusion 
coefficient \citep{Freytag1996, Herwig2000}, whose extent is also a free parameter. By 
default, the coefficient is fixed at zero (i.e., there is no mixing by convective 
overshooting). Finally, the atmosphere is included in the stellar model by placing the 
outermost mesh-point at optical depth $\tau=10^{-4}$ and applying a boundary condition 
corresponding to a gray Eddington atmosphere. This is equivalent to describing the 
atmosphere (where $\tau<2/3$) with the Eddington $T(\tau)$ relation. Gravitational 
settling of helium and metals are modeled using the prescription of \citet{Thoul1994}. 
Although gravitational settling improved the quality of standard solar models, it is 
known to erroneously predict that all metals and helium are drained from the surfaces of 
stars with masses $M\gtrsim1.2\ M_\odot$. We therefore compute gravitational settling at 
full efficiency for $M<1.1\ M_\odot$, zero efficiency for $M\geq1.2\ M_\odot$ (i.e., we 
disable it), and an efficiency that decreases linearly between $1.1\ M_\odot$ and $1.2\ 
M_\odot$.

Several details of the fitting method described in \cite{Metcalfe2009} have been updated 
for AMP~2.0. First, the range of metallicities has been slightly reduced from [0.002, 
0.05] for ASTEC models to [0.008, 0.05] for MESA. Second, we stop the evolution when the 
stellar model reaches a minimum $\log g=3.75$, to speed the computation by excluding the 
red giant phase. Third, we devised a new age interpolation scheme for the final stellar 
model. During stellar evolution we monitor the difference $y$ between the observed and 
modeled frequencies of the lowest-frequency radial mode. When $y$ changes sign, we solve 
for the root $y=0$ by applying a bisection algorithm to the numerical time-step. If 
multiple sign-changes and roots are encountered over the course of the evolution, we pick 
the one with the smallest $\chi^2$ between observed and modeled radial-mode frequencies. 
Finally, to match the observed oscillation frequencies we use the two-term correction for 
surface effects proposed by \cite{Ball2014}, rather than the empirical correction of 
\cite{Kjeldsen2008}.

\section{Deployment and Validation}

The source code for AMP~2.0 is available on 
GitHub\footnote{\url{https://github.com/travismetcalfe/amp2}}, and it is built around 
MESA release 12778 and GYRE version 6.0. Minor modifications are included for a few 
routines in the MESA source code, to minimize standard output for parallel runs. We 
deployed AMP~2.0 on the Stampede2 supercomputer at the Texas Advanced Computing Center, 
running a set of hybrid MPI-OpenMP jobs for each set of observational data. Each job is 
an MPI parallel instance of the GA, which spawns an ensemble of 120 tasks. Each task is a 
4-core OpenMP parallel instance of a stellar model produced by MESA and GYRE, so each job 
runs on 480 cores and requires about 12 hours to complete. In practice, these jobs are 
executed on 10 nodes of Stampede2 (with 48-cores each) because hyper-threading up to 
96-cores per node was found to degrade performance significantly. Each run is an ensemble 
of 4 jobs with different random initialization, so the analysis of each set of 
observational data requires 40 nodes for 12 hours (480 node-hours per run).

We validated the precision and accuracy of the stellar properties determined from AMP~2.0 
by applying it to data sets for the Sun and several solar analogs. References for the 
adopted observational constraints are given in the final column of Table~\ref{tab1}. For 
this small sample, AMP~2.0 yields a median precision of 0.8\%, 2.2\%, and 8.4\% on the 
stellar radius, mass, and age, respectively. The accuracy of AMP~2.0 can be judged from 
the inferred properties of the Sun, which are all consistent with the known solar values. 
Additional probes of the accuracy include independently determined stellar radii from 
interferometry \citep{North2007, Bazot2011, White2013, Kervella2017} and masses for the 
components of $\alpha$~Cen \citep{Kervella2016}, as well as the consistent stellar ages 
inferred for the components of $\alpha$~Cen and 16~Cyg. Luminosities inferred from 
AMP~2.0 tend to be marginally lower than measured values, which may be related to our 
choice of atmospheric boundary condition.

\begin{deluxetable*}{lccccc}[t]
\setlength{\tabcolsep}{19.5pt}
\tablecaption{Stellar Properties from the Asteroseismic Modeling Portal\label{tab1}}
\tablehead{ & \colhead{$R~(R_\odot)$} & \colhead{$M~(M_\odot)$} & \colhead{$L~(L_\odot)$} &\colhead{Age~(Gyr)} & \colhead{Data Sources}}
\startdata
Sun            & $1.001 \pm 0.005$ & $1.00 \pm 0.01$ & $0.96 \pm 0.08$ & $4.69 \pm 0.30$ & 1, 2    \\ 
18~Sco         & $0.990 \pm 0.009$ & $0.96 \pm 0.03$ & $0.95 \pm 0.07$ & $4.46 \pm 0.43$ & 3, 4, 5 \\ 
$\alpha$~Cen~A & $1.226 \pm 0.008$ & $1.12 \pm 0.02$ & $1.50 \pm 0.08$ & $6.27 \pm 0.79$ & 6, 7, 8 \\ 
$\alpha$~Cen~B & $0.864 \pm 0.007$ & $0.92 \pm 0.02$ & $0.51 \pm 0.06$ & $5.73 \pm 1.77$ & 9, 10   \\ 
16~Cyg~A       & $1.220 \pm 0.011$ & $1.06 \pm 0.03$ & $1.40 \pm 0.11$ & $7.43 \pm 0.54$ & 11, 12  \\ 
16~Cyg~B       & $1.116 \pm 0.008$ & $1.04 \pm 0.02$ & $1.21 \pm 0.08$ & $6.99 \pm 0.44$ & 11, 12  \\ 
$\beta$~Hyi    & $1.823 \pm 0.031$ & $1.10 \pm 0.05$ & $3.50 \pm 0.24$ & $6.66 \pm 0.56$ & 13      \\ 
\enddata
\tablerefs{(1)~\cite{Davies2015}; (2)~\cite{Metcalfe2015}; 
(3)~\cite{Bazot2011}; (4)~\cite{Bazot2012}; (5)~\cite{Li2012}; 
(6)~\cite{deMeulenaer2010}; (7)~\cite{Thevenin2002}; (8)~\cite{Kervella2003}; 
(9)~\cite{Kjeldsen2005}; (10)~\cite{PortodeMello2008}; 
(11)~\cite{Lund2017}; (12)~\cite{Ramirez2009};
(13)~\cite{Brandao2011}}
\end{deluxetable*}

\begin{acknowledgments}
Development of AMP~2.0 was supported by grant NNX16AB97G from the National Aeronautics 
and Space Administration (NASA). Computational time at the Texas Advanced Computing 
Center was provided through XSEDE allocation TG-AST090107. R.H.D.T.\ acknowledges NASA 
grant 80NSSC20K0515.
\end{acknowledgments}

\end{document}